\documentclass[aps,pra,10pt,superscriptaddress,numerical,floatfix,twocolumn]{revtex4-2}

\usepackage{bm}
\usepackage{braket}
\usepackage{color}
\usepackage{dcolumn}
\usepackage{epsfig}
\usepackage{float}
\usepackage[T1]{fontenc}
\usepackage{graphicx}
\usepackage[colorlinks]{hyperref}
\AtBeginDocument{
	\hypersetup{
		urlcolor=black,
		citecolor=green,
		linkcolor=blue,
	}%
}
\usepackage[latin9]{inputenc}
\usepackage{mathtools}
\usepackage{caption}
\usepackage{subcaption}
\usepackage{xcolor}
\usepackage{cleveref}
\usepackage{amssymb}
\usepackage{amsthm}
\usepackage{amsmath}

\def\app#1#2{%
	\mathrel{%
		\setbox0=\hbox{$#1\sim$}%
		\setbox2=\hbox{%
			\rlap{\hbox{$#1\propto$}}%
			\lower1.2\ht0\box0%
		}%
		\raise0.25\ht2\box2%
	}%
}

\begin{document}

\title{Entanglement of Magnetically Levitated Massive Schr\"odinger Cat States by Induced Dipole Interaction}

\newcommand{\affone}{University College London, The United Kingdom}
\newcommand{\afftwo}{Center for Fundamental Physics, Department of Physics and Astronomy, Northwestern University, Evanston, USA 60208}
\newcommand{\affthree}{Centre for Quantum Computation and Communication Technology, School of Mathematics and Physics, University of Queensland, Brisbane, Queensland 4072, Australia.}
\newcommand{\afffour}{Van Swinderen Institute, University of Groningen, 9747 AG, The Netherlands}

\author{Ryan J. Marshman}
\affiliation{\affthree}
\author{Sougato Bose}
\affiliation{\affone}
\author{Andrew Geraci}
\affiliation{\afftwo}
\author{Anupam Mazumdar}
\affiliation{\afffour}

\date{\today}

\begin{abstract}
Quantum entanglement provides a novel way to test short-distance quantum physics in a non-relativistic regime. 
We provide entanglement-based protocols to potentially test the magnetically induced dipole-dipole 
interaction and the Casimir-Polder potential between the two nano-crystals kept in a Schr\"odinger  Cat state. 
Our scheme is based on the Stern-Gerlach (SG) apparatus, where we can witness the 
entanglement mediated by these interactions for the nano-crystal mass $m\sim 10^{-19}$~kg  with a 
spatial superposition size of order $0.1$ micron in a trap relying on diamagnetic levitation. 
We show that it is possible to  close the SG interferometer in position and momentum with a 
modest gradient in the magnetic field.
\end{abstract}

\maketitle

\section{Introduction}

Quantum entanglement is a critical observable that demarcates the classical from the quantum world~\cite{RevModPhys.81.865}. Entanglement provides an unquestionable quantum signature that cannot be mimicked by any classical operation between the two quantum systems. In fact, a well-known theorem (known as local operation and classical communication (LOCC)) prohibits entangling the two quantum systems that are mediated by a classical interaction~\cite{PhysRevA.54.3824}. After all,  the Standard Model (SM) interactions are known to be quantum, and quantum entanglement-based protocols allow us to test how various forces of nature can be entangled in a laboratory setup~\cite{Barker:2022mdz}. At a short distance, but still, in the infrared (IR) regime, we can test virtual photon-induced interactions~\cite{vandeKamp:2020rqh}, such as Coulomb, Casimir-Polder~\cite{PhysRev.73.360}, and magnetically induced dipole-dipole interaction~\cite{Jackson:1998nia}. 

In search for the answer to the profound question; "does gravity follows the rules of quantum mechanics or not?" witnessing gravitationally mediated entanglement has been proposed as a key protocol to test the quantum nature of gravity in a laboratory \cite{ICTS,Bose:2017nin}(for a related work see \cite{marletto2017gravitationally}). The scheme relies on two masses, each prepared in a spatial superposition and placed at distances where they couple solely gravitationally. If gravity follows the rules of quantum interaction, and not that of a classical real-valued field, then the two masses will entangle \cite{Bose:2017nin,Belenchia:2018szb,Marshman:2019sne,Bose:2022uxe,danielson2022gravitationally,christodoulou2022locally}. However, to test the quantum nature of gravity we will need heavy masses $10^{-14}-10^{-15}$kg, and large spatial superposition of $10-100{\rm \mu m}$, with long coherence times of $1-2$ seconds. Furthermore, one can also test the quantum origin of gravitational interaction between quantum matter and light~\cite{Biswas:2022qto}, which will enable us to understand the spin nature of the gravitational interaction. All these protocols are commonly known as quantum gravity-induced entanglement of masses (QGEM)~\cite{Bose:2017nin}.
One crucial ingredient for all these experiments is to understand the entanglement generation from the known well-established photon-induced electromagnetic interactions. The photon-induced entanglement will create a background that is necessary to understand before we could perform the QGEM experiment~\cite{vandeKamp:2020rqh,2021arXiv210611906Y}. 

The aim of this paper will be to show that there exists a very natural electromagnetic background which is an inevitable consequence of any neutral nano-crystal in the presence of an external magnetic field. The external magnetic field, which we require for trapping the nano-crystal, and eventually creating the quantum superposition will induce a magnetic dipole to the nano-crystal. The two nano-crystals in a quantum superposition will get entangled by the electromagnetic interaction, and hence it will provide a dominant background for the QGEM experiment at short distances. 

The use of a Stern-Gerlach (SG) apparatus is one of the most promising approaches towards atom interferometry~\cite{bloom1962transverse,Folman2013,folman2019,Folman2018,PhysRevLett.117.143003,Margalit:2020qcy}. However, for larger objects such as nanoparticles we can also exploit the SG properties to create spatial superposition, see \cite{scala2013matter,Pedernales_2020,Marshman:2021wyk,Zhou:2022frl,Zhou:2022jug,Zhou:2022epb}.  

Such interferometers have already been realized using atom chips~\cite{folman2019,Folman2018,Margalit:2020qcy}, for both half-loop~\cite{Folman2013} and full-loop \cite{Margalit:2020qcy} configurations achieving the superposition size of $3.93$ $\mu$m and $0.38$ $\mu$m in the experimental time of 21.45 ms and 7 ms, respectively. Based on this SG scheme there have been theoretical studies to create even more ambitious superposition sizes~\cite{Zhou:2022frl,Zhou:2022jug,Zhou:2022epb}. In all these cases, the idea is to manipulate the nitrogen vacancy (NV) center of a nano-diamond. The NV center provides a spin defect, which can be manipulated in the presence of an external magnetic field. One can use the spins of two adjacent interferometers to create the entanglement witness by performing spin correlations~\cite{Bose:2017nin,PhysRevA.102.022428}.

Of course, one of the key challenges is to understand numerous sources of decoherence and noise ~\cite{Bose:2017nin,romero2010toward,PhysRevA.84.052121,romero2011large,Nguyen:2019huk,RevModPhys.85.471,vandeKamp:2020rqh,Schut:2021svd,Toros:2020dbf}. They arise from residual gas collisions and environmental photons, which can be attenuated by vacuum and low-temperature technologies~\cite{Gabrielse2016,Marx:2014ofa}. In addition,  the Humpty-Dumpty effect \cite{Schwinger1988}, internal cooling of the nanodiamond to improve  the spin coherence time~\cite{Bar-Gill(2013),Abobeih(2018),2014NatNa,PhysRevB.92.060301}, as well as to  tackle the Majorana spin-flip, under development ~\cite{Marshman:2021wyk}. Moreover, there are also a series of gravitational channels for dephasing~\cite{Toros:2020dbf,Wu:2022rdv}; the emission of gravitons is negligible~\cite{Toros:2020krn}, gravity gradient noise (GGN) can be mitigated with an exclusion zone~\cite{Toros:2020dbf}, and relative acceleration can be mitigated by improving the vacuum and isolating the experimental box as much as possible.

In this work, we will aim to understand two particular EM-induced potentials which  will inevitably entangle any two quantum systems, even if we assume that the monopole contribution is neutralized, which is experimentally feasible~\cite{frimmer2017controlling}. The two potentials are the Casimir-Polder (CP) potential and the dipole-dipole (DD)  potential due to induced-dipole moment in the presence of an external magnetic field, present in the SG setup. In this paper, we will provide a superposition scheme in a levitating setup to witness the entanglements due to these potentials. We will show that with a modest magnetic field gradient in the SG setup, we can measure the entanglement witness for both CP and DD potentials for a mass of neutral nano-diamond $m\sim 10^{-19}{\rm Kg}$, and the spatial superposition of $0.1{\rm \mu m}$.


A diamagnetic material in a magnetic trapping potential will evolve according to the Hamiltonian 
\begin{equation}
	\hat{H}=\frac{\hat{\bf{p}}^2}{2m}+\hbar D \hat{S}_z+g_s\frac{e\hbar}{2m_e}\hat{\bf{S}} \cdot{\bf{B}}+mg\hat{y}-\frac{\chi_{\rho} m}{2\mu_0}{\bf{B}^2} \label{eq:hamiltonian}
\end{equation}
where the first term in Eq.(\ref{eq:hamiltonian}) represents the kinetic energy of the nanodiamond, $\hat{\boldsymbol{p}}$ is the momentum operator and $m$ is the mass of the nanodiamond. The second term represents the zero-field splitting of the NV center with $D=(2\pi)\times2.8~{\rm GHz}$, $\hbar$ is the reduced Planck constant, and $\hat{S}_{z'}$ is the spin component operator aligned with the NV axis. The third term represents the interaction energy of the NV electron spin magnetic moment with the magnetic field  $\boldsymbol{B}$. The spin magnetic moment operator $\hat{\boldsymbol{\mu}}=-g_{s}\mu_{B}\hat{\boldsymbol{S}}$, where $g_{s}\approx 2$ is the Land\`{e} g-factor, $\mu_{B}=e \hbar/2 m_{e}$ is the Bohr magneton and $\hat{\boldsymbol{S}}$ is the NV spin operator. The fourth term is the gravitational potential energy, $g\approx9.8$ ${\rm m/s^{2}}$ is the gravitational acceleration and $\hat{\boldsymbol{z}}$ is the position operator along the direction of gravity ($z$ axis). The final term represents the magnetic energy of a diamagnetic material (diamond) in a magnetic field, $\chi_{\rho}=-6.2\times10^{-9}$ ${\rm m^{3}/kg}$ is the mass susceptibility and $\mu_{0}$ is the vacuum permeability. 

For the purpose of our scheme, we will make use of a well-known trap profile ${\bf{B}}_T$ given by \cite{hsu2016cooling}
\begin{widetext}\label{Trapping potential}
	\begin{align}
		{\bf{B}}_T=&-\left[\frac{3a_4\sqrt{\frac{35}{\pi}}x^2y}{8 y_0^3} +\frac{3a_4\sqrt{\frac{35}{\pi}}y\left(x^2-y^2\right)}{16 y_0^3} -\frac{a_3\sqrt{\frac{7}{6\pi}}x^2}{y_0^2} +\frac{a_2\sqrt{\frac{15}{\pi}}y}{4 y_0} +\frac{a_3\sqrt{\frac{7}{6\pi}}\left(-x^2-y^2+4z^2\right)}{2 y_0^2}\right]\hat{x} \nonumber\\
		&-\left[-\frac{3a_4\sqrt{\frac{35}{\pi}}xy^2}{8 y_0^3} +\frac{3a_4\sqrt{\frac{35}{\pi}}x\left(x^2-y^2\right)}{16 y_0^3} -\frac{a_3\sqrt{\frac{7}{6\pi}}xy}{y_0^2} +\frac{a_2\sqrt{\frac{15}{\pi}}x}{4 y_0} \right]\hat{y}-\left[\frac{2a_3\sqrt{\frac{14}{3\pi}}xz}{y_0^2}\right]\hat{z} \label{eq:Btrap}
	\end{align}
\end{widetext}
where $y_0=75~\mu$m is the distance from the center of the trap to the pole pieces which help generate the trap and $a_2=-1.3$ T, $a_3=0.0183$ T, and $a_4=0.72$ T determine the magnetic field strength.
We will take the particle at time $t=0$ to be at rest in the trapping potential, with an initial spin state $\frac{1}{\sqrt{2}}\left(\left|+1\right\rangle+\left|-1\right\rangle\right)$. 
The trapping potential is given by $U=(\chi_\rho m B^2/2\mu_0)+mgy$, since $\chi<0$, the particle can be trapped at a frequency 
$\omega_{\zeta}=\sqrt{-(\chi/2\rho\mu_0)(\partial^2B^2/\partial\zeta^2)}$, where $\zeta=x,y,z$.

A large gradient, linear magnetic field $\vec{B}_ P$ is then pulsed on, for a short time $t_p$ which acts to create a momentum difference correlated with the internal spin state of the particle. The particle then evolves in the weakly trapping potential for some time $t_T$ before the large gradient linear magnetic field is again turned on for a further time $t_P$. At this point, the particle should be returned to its initial position at time $t=2t_{P}+t_T\equiv T$, with the internal spin state now in the form $\frac{1}{\sqrt{2}}\left(e^{i\phi_+}\left|+1\right\rangle+e^{-\phi_-}\left|-1\right\rangle\right)$ where the phases $\phi_{\pm}$ are a result of the interactions with external sources.

To create the superposition we will first assume the pulsed magnetic field is linear over the region experienced by the particle, with a profile
\begin{equation}
	{\bf{B}}_p=\eta\left(y(t=0)-y\right)\hat{y}+\eta z\hat{z} \label{eq:Bpulse}
\end{equation}
where $\eta$ gives the magnetic field gradient and $y(t=0)$ is the initial position of the particle in the $y$ direction, which is away from the minimum of the potential for the particle. We wish to start high up in the potential to gain large momentum, which will help the two spin states to get separated in $z$-direction as much as possible. With this definition of the pulse field, it makes a minimal disruption to the trapping field, being zero magnitudes (although non-zero gradient) at the particle's location ($x=0$, $y\approx y(t=0)$, and $z\approx 0$). 

\begin{figure*}
	\centering
    \begin{subfigure}[l]{0.47\columnwidth}
		\includegraphics[width=\columnwidth]{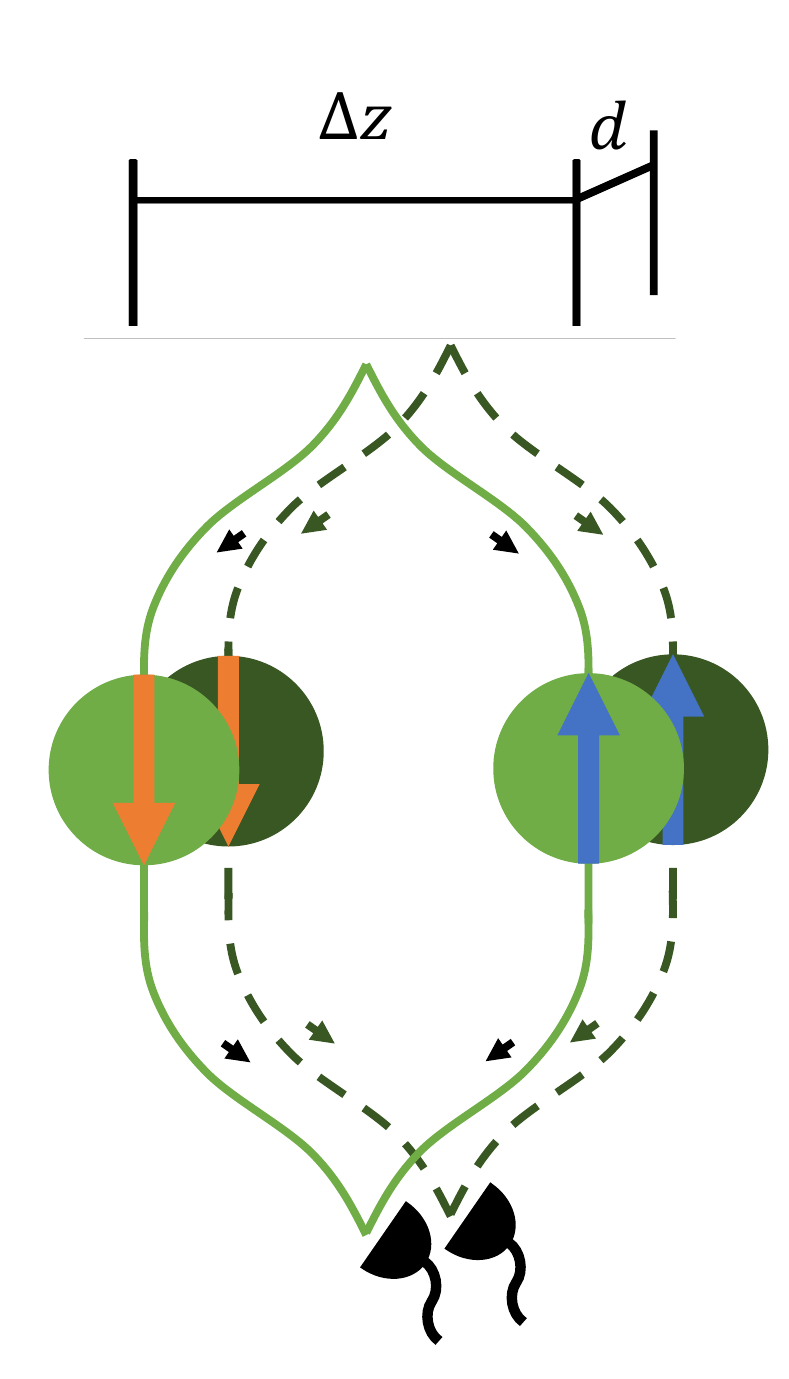}
		\caption{Simplified interferometer arrangement. \label{fig:system diagram}}
	\end{subfigure}
	\begin{subfigure}[l]{0.8\columnwidth}
		\includegraphics[width=\columnwidth]{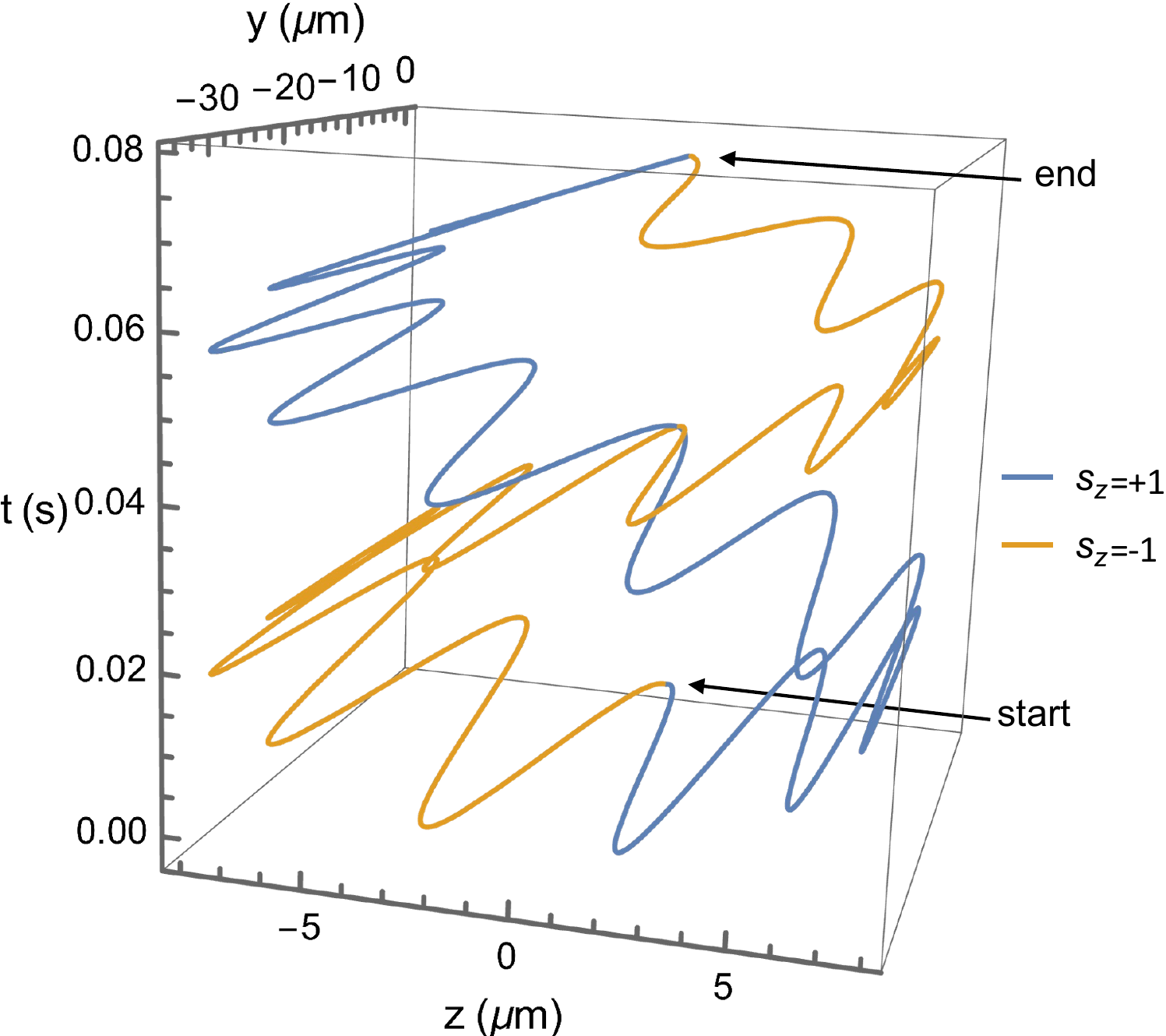}
		\caption{Space-time trajectory of single interferometer.\label{fig:space-time trajectory}}
	\end{subfigure}\\
 	\begin{subfigure}[l]{0.95\columnwidth}
		\includegraphics[width=\columnwidth]{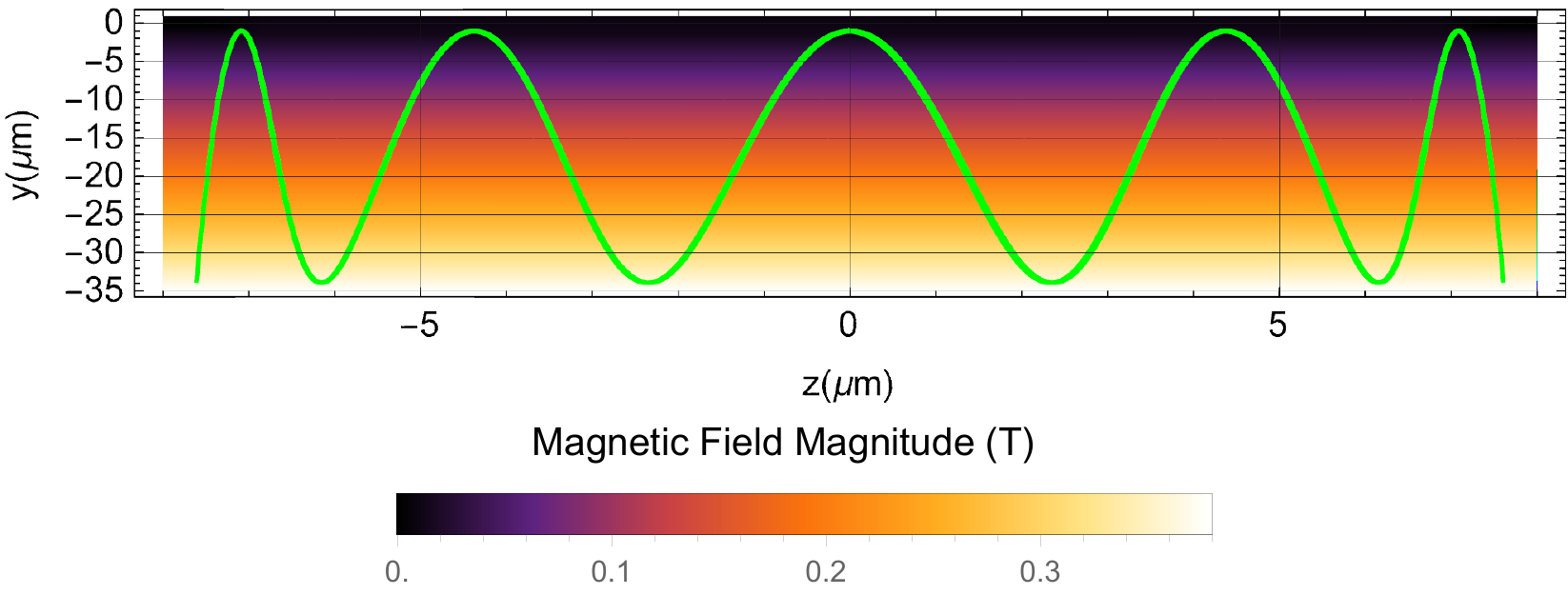}
		\caption{Trajectory in the $z-y$ plane.\label{fig:trajectory through magnetic field}}
	\end{subfigure}
 	\begin{subfigure}[l]{0.8\columnwidth}
		\includegraphics[width=\columnwidth]{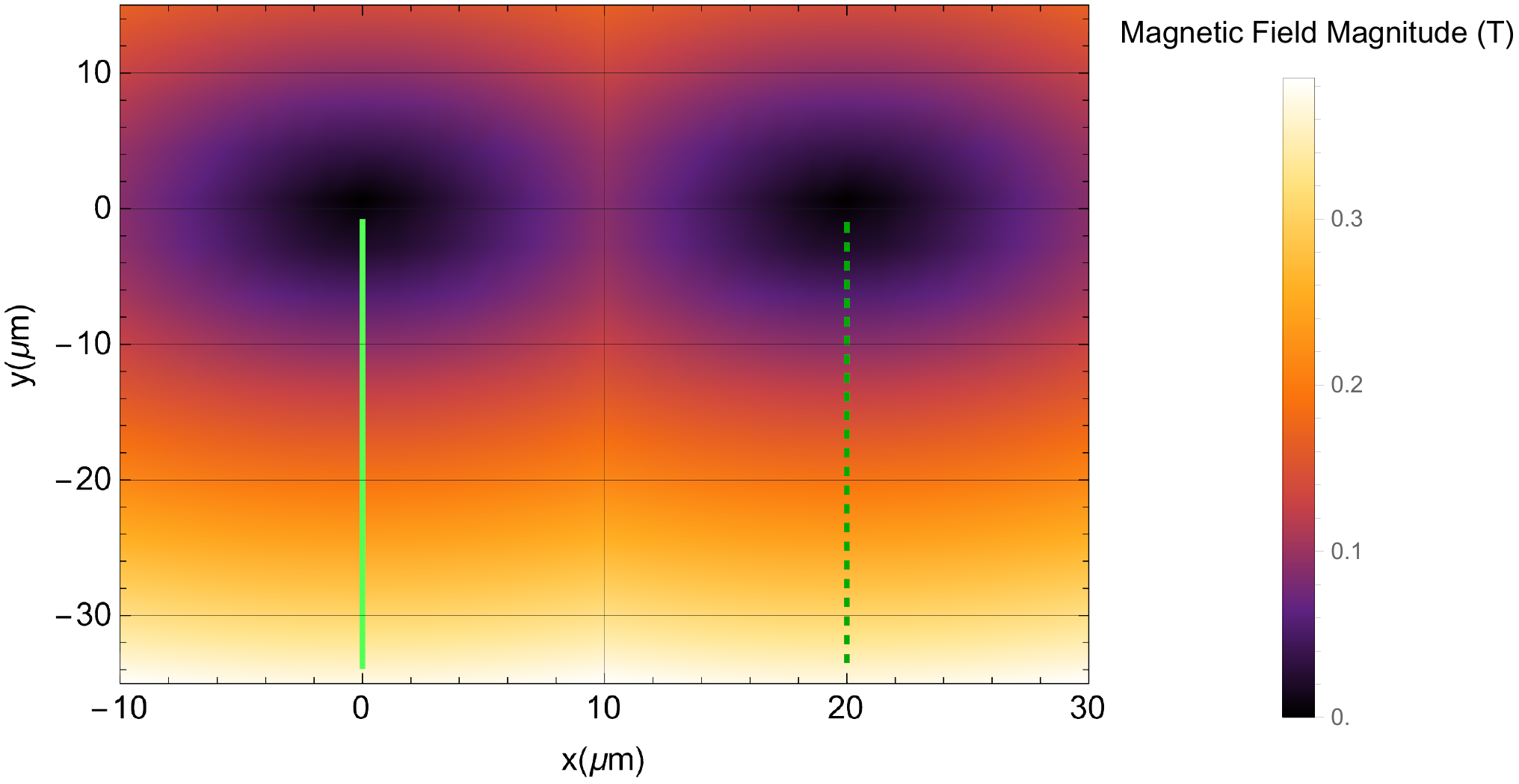}
		\caption{Trajectory in the $x-y$ plane.\label{fig:xy-trajectory through magnetic field}}
	\end{subfigure}
	\caption{\ref{fig:system diagram} shows a schematic representation of the manner in which two of these devices can be arranged, offset by a distance $d$ in the $x$ direction to enable the generation of entanglement between the two particles.\ref{fig:space-time trajectory} shows and actual space-time trajectory for a two $z$ oscillation interferometer. \ref{fig:trajectory through magnetic field} shows the same trajectory in the $y-z$ plane and the trapping magnetic field while \ref{fig:xy-trajectory through magnetic field} shows the same trajectory in the $x-y$ plane, for an interferometer distance $d=20$ $\mu$m. All trajectories are for $m=3.8\times10^{-19}$ kg masses, a magnetic field gradient pulse time $t_p\approx160~\mu$s and the particle is started at $y(t=0)=-1.11~\mu$m.  \label{fig:trajectories}}
\end{figure*}

We can control the average magnitude of the diamagnetically induced magnetic field of the diamond by initially displacing the particle in the $y$ direction. This will result in the diamond oscillating in the $y$ direction according to the trapping frequency in this direction. This trapping frequency in the $y$ direction will be much faster than that in the $z$ (splitting) direction as the particle is intended to be more confined in this direction. This oscillation is seen in Figure \ref{fig:space-time trajectory} which shows an example space-time trajectory for each arm of the interferometer if it is allowed to oscillate for a full period before the pulsed magnetic field is re-applied to close the spatial superposition. The dual oscillatory behavior is due to the mass not occupying the ground state of the trapping potentials in both $y$ and $z$ directions, with the $y$ trapping frequency in the $z$ direction, i.e. $\omega_y\ll\omega_z$.
To ensure that both the spatial wavefunction completely matches at time $t=T$, it is necessary to ensure the trapping frequency in the $y$ direction is an integer multiple of that in the $z$ direction such that both the position and momentum matches at the final time. It will also be necessary to ensure the pulsed magnetic field opening and closing the superposition matches in both magnitude and pulse time, with deviations from this leading to reduced contrast in the form of decoherence in the final spin states. 

The pulse time $t_p= 2\pi/\omega$ is determined by the pulsed magnetic field gradient, $\eta$, to ensure the maximum momentum difference is induced between the two masses, we are interested in a quarter of the oscillation period or $t_p/4=\pi/(2\omega)$, where the frequency is given by the diamagnetic-induced magnetic field $\omega = \sqrt{k/m}$, where $k$ is the spring constant determined by the effective potential of the crystal, $U_{\pm}(y)=-(\chi_\rho m/2\mu_0)\eta^2y^2$ 
\begin{equation}
	t_p=\frac{\pi}{2\eta}\sqrt{\frac{-\mu_0}{2\chi_{\rho}}}
\end{equation}
This time is a quarter of the oscillation period of the Harmonic trap caused by the pulsed magnetic field

All results were numerical simulations of the classical equations of motion derived from the Hamiltonian given in Eq. \ref{eq:hamiltonian}. The magnetic profile used was ${\bf{B}}={\bf{B}}_T+{\bf{B}}_p(t)$ where ${\bf{B}}_T$ is given by Eq. \ref{eq:Btrap} and ${\bf{B}}_p(t)=0$ any time the linear gradient pulse is switched off, otherwise it is given by Eq. \ref{eq:Bpulse}. Thus, the pulsed magnetic field was taken to be switched on and off instantly.

Note that, as shown in Figure \ref{fig:trajectory through magnetic field}, due to the downward shift of the rest position of the mass in the background magnetic field due to the downward pull of gravity, the zero-field region of the magnetic field can be avoided by simply initializing the mass away from it. This serves the dual purpose of generating the induced diamagnetic field in the particle and naturally avoiding Majorana spin flips.

%

Once, we have created one superposition in a trapping potential given by Eq.~(\ref{Trapping potential}), we can imagine bringing two such trapping potentials, hence bringing two interferometers close to each other at a distance $d$ shown in Fig.~(\ref{fig:system diagram}), separated by a distance $d$. This is the parallel setup of the original QGEM proposal, studied in \cite{Nguyen:2019huk,Barker:2022mdz,Tilly:2021qef}. 

If two such interferometers are placed near one another, separated in the $x$ direction by a distance $d$ with the direction the spatial splitting parallel as shown in Figure \ref{fig:system diagram}, we expect the joint state to evolve to
\begin{align}
	\left|\Psi\right\rangle=&\frac{1}{2}\left(\left|+1,+1\right\rangle+\left|-1,-1\right\rangle\right) \nonumber\\
	&+\frac{e^{i\Delta\phi}}{2}\left(\left|+1,-1\right\rangle+\left|-1,+1\right\rangle\right)
\end{align}
where $\Delta\phi=\phi_{+-}-\phi_{++}$ and $\phi_{ij}$ is the phase due to the interaction between the $i$ and $j$ arms of the interferometer. We refer to this phase difference $\Delta\phi$ as the entanglement phase, which is actually the summation of all particle-particle interactions, as considered separately in the following section.

Now, we are able to experimentally probe many interesting questions, such as various particle-particle interactions mediated by photons, which may be witnessed via the entanglement phase. These may include  Casimir-Polder (cp), and induced dipole-dipole (dd), assuming that our crystal is charged neutral (experimentally feasible~\cite{frimmer2017controlling}).  Depending on the chosen experimental parameters, it's likely that at least one of these interactions will be negligible, indeed, we will consider a situation where only the dd interaction produces a significant effect, with the cp interaction serving as the primary background. 
The potentials for each interaction are given by
\begin{align}
	U_{cp}(d)=&-\frac{23\hbar c}{4\pi}\frac{\varepsilon-1}{\varepsilon+2}\frac{\left(\frac{3m}{4\pi\rho}\right)^2}{d^7}\\
	U_{dd}(d)=&\frac{2\chi_{\rho}^2 m^2\left|\vec{B}(x)\right|^2}{4\pi\mu_0d^3}
\end{align}
where $m,~\rho,\chi_{\rho}$ and $\varepsilon$ are the particle's  mass, density, and magnetic mass susceptibility respectively and $d$ is the center of the mass distance between the particles. Note that we consider the magnetic field across the particles to be approximately constant and given by the magnetic field at the centre of mass. These interactions $i$ will each give rise to an entanglement phase given by
\begin{equation}
	\Delta\phi_{i}=\frac{1}{\hbar}\int_{0}^{T}\textrm{d}t~U_{i}\left(d_1(t)\right)-U_{i}\left(d_2(t)\right)
\end{equation}
where $d_1(t)$ ($d_2(t)$) is the furthest (closest) separation distance between the two particles due to the two superposed spatial states.

\begin{figure*}
	\centering
	\begin{subfigure}[l]{0.9\columnwidth}
		\includegraphics[width=\columnwidth]{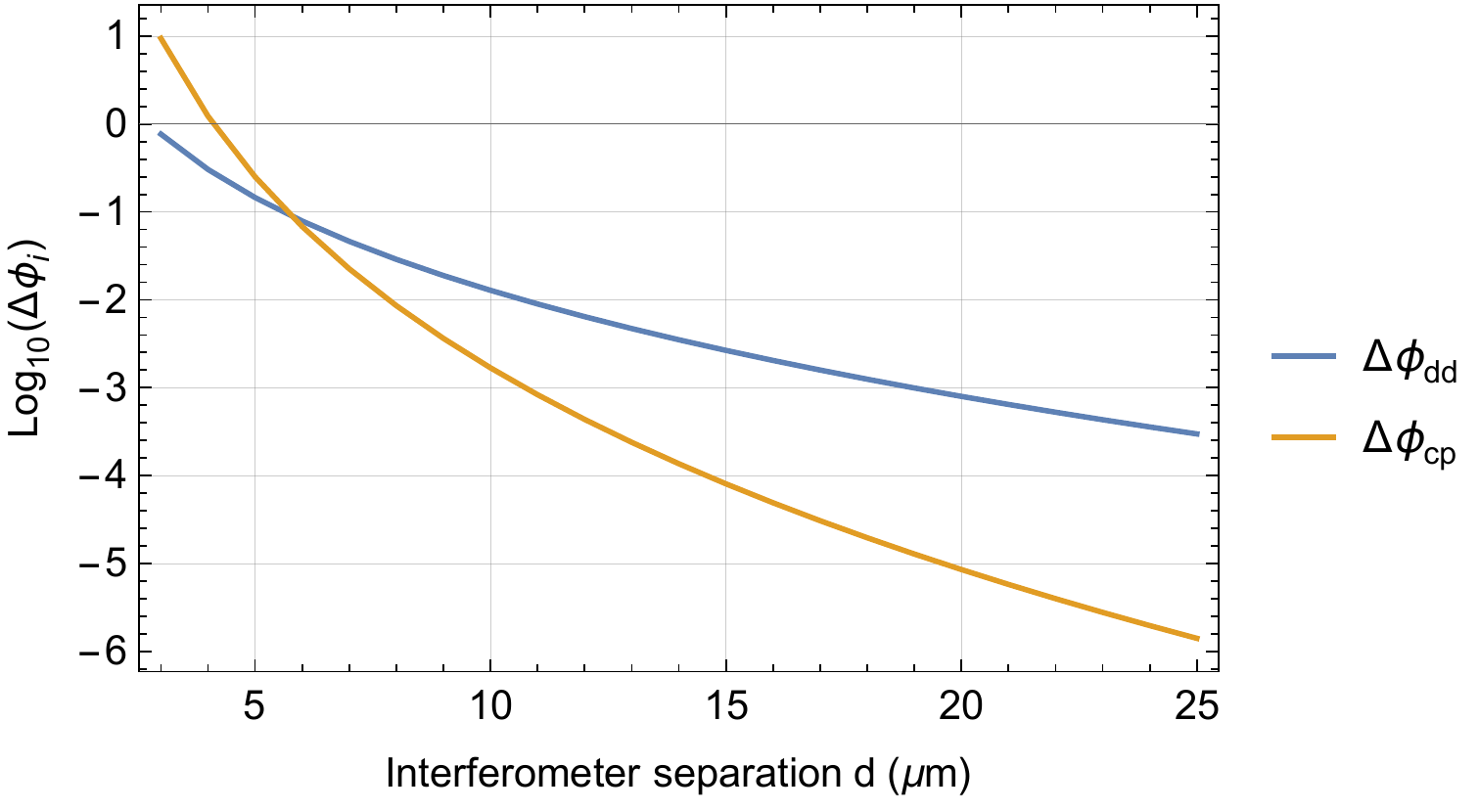}
		\caption{Entanglement phase comparison for different sources. \label{fig:entanglement phase}}
	\end{subfigure}
	\begin{subfigure}[l]{1.05\columnwidth}
		\includegraphics[width=\columnwidth]{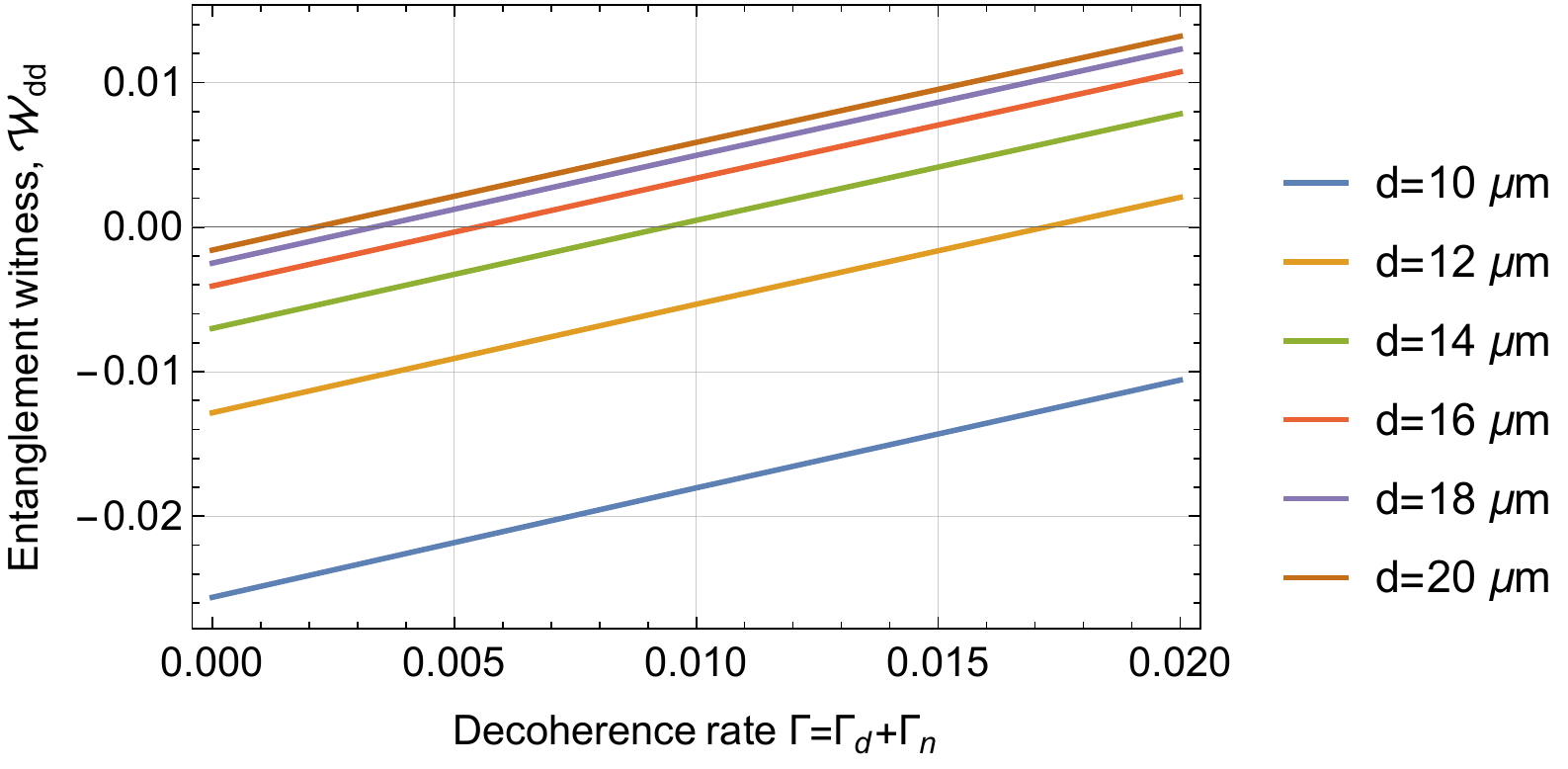}
		\caption{Entanglement witness value as a function of total decoherence $\Gamma$ for various interferometer separation distances $d$.\label{fig:entanglement witness}}
	\end{subfigure}
	\caption{Entanglement phase and witness values comparisons between different signal sources as a function of the total decoherent $\Gamma$. Here each interferometer is as shown in Figure \ref{fig:space-time trajectory}, with $m=3.8\times10^{-19}$ kg, $T\approx0.077$ s, $t_p\approx160~\mu$s which gave the particles enough time to complete one full period of oscillation in the $z$ direction. }
	\label{fig:entanglement phase and witness}
\end{figure*}

The entanglement can then be found by measuring an entanglement witness, such as see~\cite{torovs2021relative}
\begin{equation}
	\mathcal{W}_i=1-\left(2e^{-\frac{1}{2}\left(\Gamma_n+\Gamma_d\right)}\sin\left(\Delta\phi_i\right)+\frac{1}{2}\left(e^{-2\Gamma_n-\Gamma_d}+1\right)\right)
\end{equation}
where $\Gamma_d$ and $\Gamma_n$ are the damping and noise decoherence respectively \cite{torovs2021relative}. Note that as defined here, this decoherence is defined as the decoherence rate multiplied by interferometer time. We plot the witness ${\cal W}_{dd}$ with regards to total decoherence and damping rate: $\Gamma=\Gamma_d+\Gamma_n$, see Fig.~(\ref{fig:entanglement witness}). To witness the entanglement, we require ${\cal W} < 0$~\cite{PhysRevA.102.022428}.

Fig. \ref{fig:entanglement phase} shows how the entanglement phase from each particle-particle differs with regards to the separation between the two matter-wave interferometers for both the external magnetic field-induced dipole-dipole entanglement and the CP-induced entanglement.  We can see that the entanglement witness due to the magnetic dipole-dipole interaction dominates, as expected, at distances $d> 6{\rm \mu m}$, see~Fig. \ref{fig:entanglement witness}. Below this CP-induced entanglement dominates, as expected again due to the fall-off of the interaction strength with regard to the separation distance $d$. In Fig. \ref{fig:entanglement witness}, we have taken the range of decoherence rates and shown that the dominant entanglement is due to the induced magnetic dipole-dipole interaction. Here we have assumed the mass is similar in both the interferometers, $m=3.8\times 10^{-19}$kg, and the largest superposition we are generating around $\sim 1.1{\rm \mu m}$.

Before we conclude, we should mention that the time duration of coherence for the NV spin is one of the limiting factors, but the spin coherence times are perpetually increasing (approaching 1s \cite{Bar-Gill(2013)}, even 30 s \cite{Abobeih(2018)}), but adapting to our scenario remains an open challenge~\cite{PhysRevB.105.205401}. We will also have to make sure that the external temperature is below $1$k and the gas pressure is near ${\cal O}(10^{-15})$ Pa. In our analysis we have assumed that the NV spin is not wobbling due to external torque, in reality, the NV spin will processes, and future analysis has to be taken into account~\cite{Japha:2022xyg}. The nano-diamond can also wobble in the presence of the external magnetic field~\cite{Stickler_2021}. However, this can be suppressed even before being released from the trap, for example, by using anisotropically shaped nanoparticles, which can be aligned with any given direction in space by using linearly polarized lasers or electric fields~\cite{stickler2021quantum} or magnetic fields \cite{perdriat2021angle}. 
Besides, there will be vibrational excitations from the breathing mode of the nano-diamond, e.g. phonon vibration, which we will need to analyze for this experiment. The phonon vibration can be suppressed for any state manipulation which does not excite the phonons in resonance. We also note that it has already been shown that the internal degrees of freedom (phonons) do not pose a problem \cite{henkel2021internal}. 

To conclude, we have found a new entanglement  for any matter-wave interferometers, which relies on creating the macroscopic quantum superposition with the help of a neutral diamagnetic nano-object in the presence of an external magnetic field. Here we have shown for the first time an explicit scheme to create a small-scale spatial superposition to test the known electromagnetic-induced entanglements, the dominant effect arises from external magnetic field-induced dipole-dipole entanglement which dominates for the external magnetic field of ${\cal O}(1){\rm T}$, with a separation of ${\cal O}(10){\rm \mu m}$ for a mass of order $10^{-19}$kg.

As we have shown in Figs.~\ref{fig:entanglement phase},\ref{fig:entanglement witness}, the entanglement is generated via the external magnetic field induced in the nano-diamond, which is essential to create the trapping potential and the superposition.  The witness depends on the total decoherence rate and the separation distance $d$. Such a system also allows the characterization of what would be background effects in gravitational mediated entanglement experiments. Indeed, we have shown how optimizing the distance and trajectories can have a profound effect on the various entangling interactions.

\section*{Acknowledgements}
 R.J.M. is supported by the Australian Research Council (ARC) under the Centre of Excellence for Quantum Computation and Communication Technology (CE170100012). A.G. is supported in part by NSF grants PHY-2110524 and PHY-2111544, the Heising-Simons Foundation, the John Templeton Foundation, the W. M. Keck foundation, and ONR Grant N00014-18-1-2370. S.B. would like to acknowledge EPSRC grants (EP/N031105/1, EP/S000267/1 and EP/X009467/1) and grant ST/W006227/1.

\bibliography{bibliography}

\appendix
\section{Equations of Motion}
A diamagnetic material in a magnetic trapping potential will evolve according to the Hamiltonian 
\begin{equation}
	\hat{H}=\frac{\hat{\bf{p}}^2}{2m}+\hbar D \hat{S}_z+g_s\frac{e\hbar}{2m_e}\hat{\bf{S}} \cdot{\bf{B}}+mg\hat{y}-\frac{\chi_{\rho} m}{2\mu_0}{\bf{B}^2} \label{eq:hamiltonian}
\end{equation}
where the first term in Eq.(\ref{eq:hamiltonian}) represents the kinetic energy of the nanodiamond, $\hat{\boldsymbol{p}}$ is the momentum operator and $m$ is the mass of the nanodiamond. The second term represents the zero-field splitting of the NV centre with $D=(2\pi)\times2.8~{\rm GHz}$, $\hbar$ is the reduced Planck constant and $\hat{S}_{z'}$ is the spin component operator aligned with the NV axis. The third term represents the interaction energy of the NV electron spin magnetic moment with the magnetic field  $\boldsymbol{B}$. Spin magnetic moment operator $\hat{\boldsymbol{\mu}}=-g_{s}\mu_{B}\hat{\boldsymbol{S}}$, where $g_{s}\approx 2$ is the Land\`{e} g-factor, $\mu_{B}=e \hbar/2 m_{e}$ is the Bohr magneton and $\hat{\boldsymbol{S}}$ is the NV spin operator. The fourth term is the gravitational potential energy, $g\approx9.8$ ${\rm m/s^{2}}$ is the gravitational acceleration and $\hat{\boldsymbol{z}}$ is the position operator along the direction of gravity ($z$ axis). The final term represents the magnetic energy of a diamagnetic material (diamond) in a magnetic field, $\chi_{\rho}=-6.2\times10^{-9}$ ${\rm m^{3}/kg}$ is the mass susceptibility and $\mu_{0}$ is the vacuum permeability. 

\begin{figure}
	\centering
	\includegraphics[width=\columnwidth]{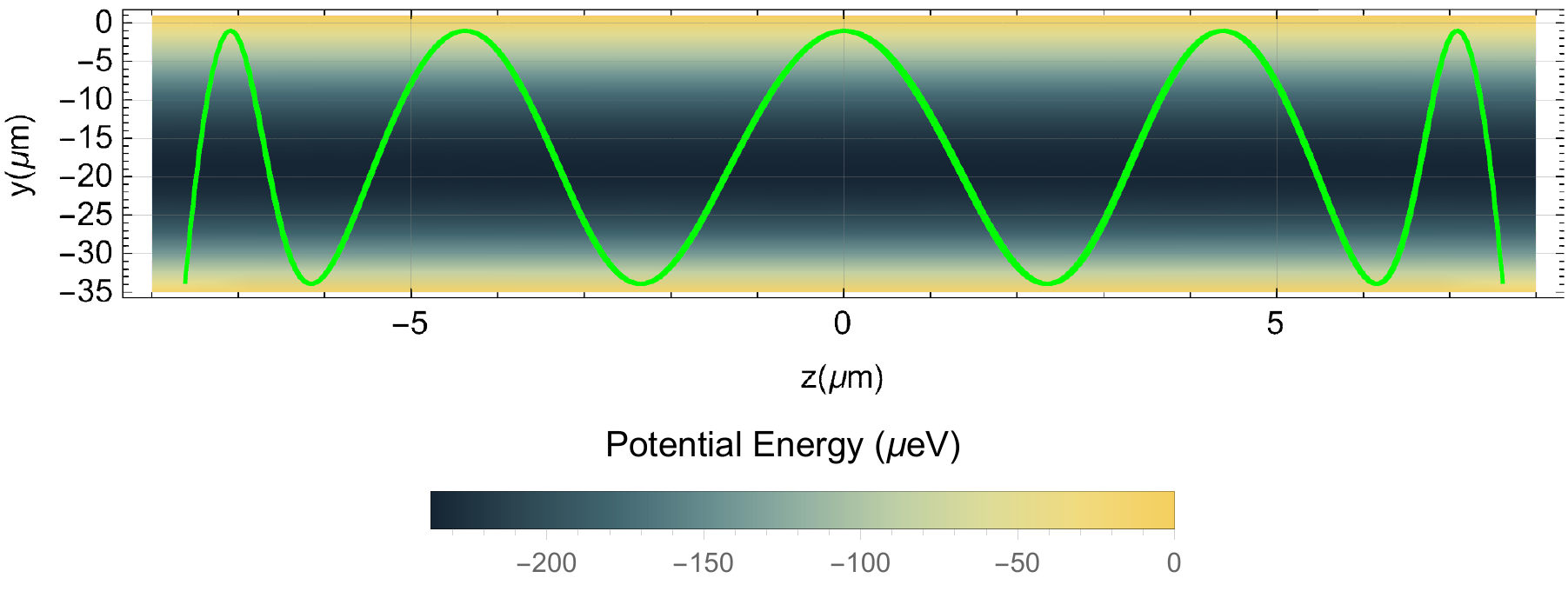}
	\caption{Example interferometer trajectory superimposed on the potential energy due to the combination of gravity and the trapping magnetic field in the $y-z$ plane. \label{fig:trajectory-through-potential}}
\end{figure}

With this we can write the equations of motion as
\begin{align}
	\partial^2_{t}{y}=& \frac{\chi_{\rho}}{2\mu_0}\partial_y{\bf{B}}^2-g_s\frac{e\hbar}{2mm_e}S_y\partial_yB_y-mg\\
	\partial^2_{t}{z}=& \frac{\chi_{\rho}}{2\mu_0}\partial_y{\bf{B}}^2-g_s\frac{e\hbar}{2mm_e}S_z\partial_zB_z-mg
\end{align}
where we're considering the motion in only 2 dimensions and the form of ${\bf{B}}$ is modified by the presence or absence of the pulsed magnetic field. The total trajectory was then solved numerically and piece wise depending on whether the pulsed magnetic field is present. The numeric trajectory is shown with the potential energy due to the trapping potential is shown in Figure \ref{fig:trajectory-through-potential}

\begin{figure}
	\centering
	\includegraphics[width=\columnwidth]{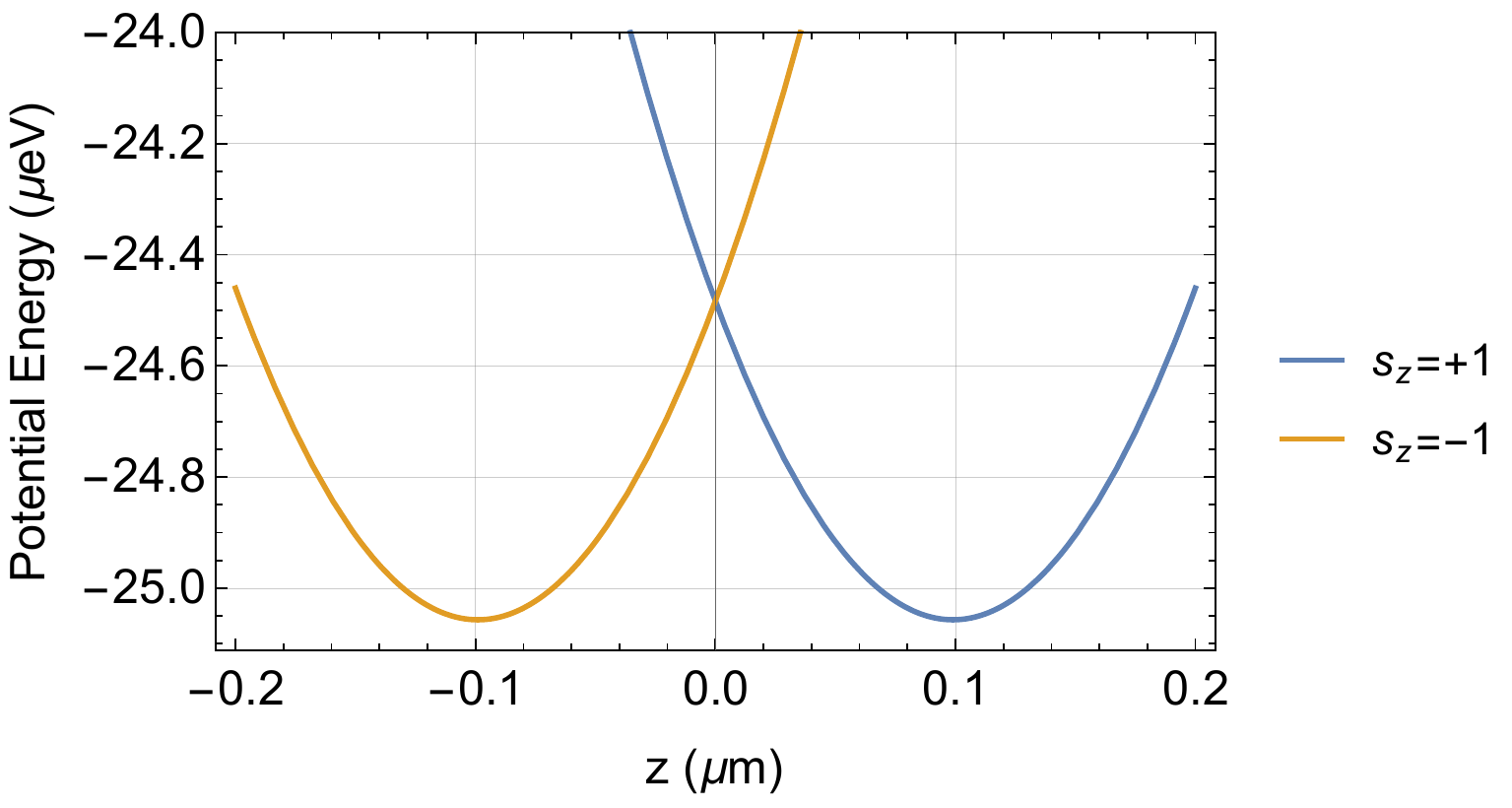}
	\caption{Potential energy for each spin state during pulsed, linear magnetic field gradient. This is shown here in the starting $x=0$, $y=-1.111$ $\mu$m plane, the pulsed magnetic field is switched off as the particles reach the minimas at $z\approx\pm0.1$ $\mu$m. \label{fig:pulsed potential}}
\end{figure}

As a reminder, the magnetic field used for the trap is given by \cite{hsu2016cooling}
\begin{widetext}\label{Trapping potential}
	\begin{align}
		{\bf{B}}_T=&-\left[\frac{3a_4\sqrt{\frac{35}{\pi}}x^2y}{8 y_0^3} +\frac{3a_4\sqrt{\frac{35}{\pi}}y\left(x^2-y^2\right)}{16 y_0^3} -\frac{a_3\sqrt{\frac{7}{6\pi}}x^2}{y_0^2} +\frac{a_2\sqrt{\frac{15}{\pi}}y}{4 y_0} +\frac{a_3\sqrt{\frac{7}{6\pi}}\left(-x^2-y^2+4z^2\right)}{2 y_0^2}\right]\hat{x} \nonumber\\
		&-\left[-\frac{3a_4\sqrt{\frac{35}{\pi}}xy^2}{8 y_0^3} +\frac{3a_4\sqrt{\frac{35}{\pi}}x\left(x^2-y^2\right)}{16 y_0^3} -\frac{a_3\sqrt{\frac{7}{6\pi}}xy}{y_0^2} +\frac{a_2\sqrt{\frac{15}{\pi}}x}{4 y_0} \right]\hat{y}-\left[\frac{2a_3\sqrt{\frac{14}{3\pi}}xz}{y_0^2}\right]\hat{z} \label{eq:Btrap}
	\end{align}
\end{widetext}
where $y_0=75~\mu$m is the distance from the center of the trap to the pole pieces which help generate the trap and $a_2=-1.3$ T, $a_3=0.0171$ T, and $a_4=0.72$ T determine the magnetic field strength. The linear, pulsed magnetic field used was
\begin{equation}
	{\bf{B}}_p=\eta\left(y(t=0)-y\right)\hat{y}+\eta z\hat{z} \label{eq:Bpulse}
\end{equation}
where $\eta$ gives the magnetic field gradient and $y(t=0)$ is the initial position of the particle in the $y$ direction, which is away from the minimum of the potential for the particle. The potential energy as experienced by each spin state during the pulsed magnetic field is shown in Figure \ref{fig:pulsed potential}.
\end{document}